\documentclass{ws-procs975x65}

\begin{document}

\title{On the thermal boundary condition of the wave function of the Universe\footnote{This research work was supported by the grants POCI/FP/63916/2005,FEDER-POCI/P/FIS/57547/2004 and Ac\c c\~oes Integradas
(CRUP-CSIC) Luso-Espanholas E-138/04.}}

\author{Mariam Bouhmadi-L\'{o}pez$^{1,2}$\footnote{mariam.bouhmadi@fisica.ist.utl.pt} and Paulo Vargas Moniz$^{2,1}$\footnote{pmoniz@ubi.pt}}

\address{$^{1}$Centro Multidisciplinar de Astrof\'{\i}sica - CENTRA,
 Departamento de F\'{\i}sica,\\ Instituto Superior T\'ecnico,
Av. Rovisco Pais 1, 1096 Lisboa, Portugal\\
$^{2}$Departamento de F\'{\i}sica,
Universidade da Beira Interior,\\ Rua Marqu\^{e}s d'Avila e Bolama,
6200 Covilh\~{a}, Portugal}

\vspace{-0.3cm}
\begin{abstract}
We broaden the domain of application of the 
recently proposed thermal boundary condition of the wave function of the Universe, which has been suggested as the basis of a dynamical selection principle on the landscape of string solutions. 
\end{abstract}

\vspace{-0.1cm}
\section{Introduction}\label{intro}

The existence of a  multiverse of vast solutions 
\cite{Bousso:2000xa,Douglas:2003um,Susskind:2003kw} 
to string theory  
constitutes 
currently an important challenge: 
How to  select a Universe or a class from the multiverse  
that will bear significant similarities  to ours?

The framework of quantum cosmology 
\cite{Vilenkin:1983xq,Hartle:1983ai,Linde:1983cm} 
provides a methodology  to establish a probability distribution for the 
dynamical parameters of the Universe. 
In this context, Brustein and de Alwis  proposed in \cite{Brustein:2005yn}, using FRW quantum cosmology,  a  
dynamical selection principle\footnote{See also Refs.~\cite{Sarangi:2005cs,Holman:2005eu}.}  on the 
landscape of string solutions \cite{Susskind:2003kw}. 

We prove that the thermal boundary condition applied in  
\cite{Brustein:2005yn} corresponds to the particular physical situation 
where the amount of radiation is very large. We then provide a broader
and improved analysis of the {\em generalised thermal boundary condition} that is independent of such  restrictive limit \cite{BLVM2006}; i.e. we consider 
an arbitrary amount of radiation consistent with the tunnelling of a closed radiation-filled Universe with a positive cosmological constant.

\enlargethispage*{8pt}
\section{The generalised thermal boundary condition}

The  thermal boundary condition for the wave function of the Universe  states that  the Universe emerges from the string era in a thermally 
excited state above the Hartle-Hawking (HH) vacuum \cite{Brustein:2005yn}.
Furthermore, the primordial thermal bath is   effectively described 
by a radiation fluid  whose energy density $\rho$ ``depends'' on the cosmological constant $\lambda$ 
\begin{eqnarray}
\rho=\frac{3\tilde{K}}{8\pi\textrm{G}}\frac{1}{a^4}, \quad
\tilde{K}\simeq \frac{\nu}{\lambda^2}. \label{K1}
\end{eqnarray}
In the previous expressions $\tilde{K}$ and $\nu$ are parameters that quantify the amount of radiation, $\textrm{G}$ is the gravitational constant and $a$ the scale factor.  Therefore, the transition amplitude of a closed radiation-filled FRW Universe\cite{Vilenkin:1998rp,Bouhmadi-Lopez:2002qz} to tunnel from the first Lorentzian region ($a<a_-$, see Fig.~1-a) to the larger Lorentzian region ($a_+<a$, see also Fig.~1-a)  within a WKB approximation reads \cite{Brustein:2005yn,BLVM2006}

\begin{equation}
\mathcal{A}=\exp(\epsilon 2I), \quad
I=\frac{\pi}{2^{\frac32}\textrm{G}}\frac{1}{\nu}g,
\label{transition2}
\end{equation}
where $\epsilon=\pm 1$ and
\begin{eqnarray}
g=\frac{\nu}{\lambda}{\sqrt{1+m}}\left[\textrm{E}(\alpha_{II})-(1-m)\textrm{K}(\alpha_{II})\right],\,\,
\alpha_{II}=\sqrt{\frac{2m}{1+m}}, \,\, m=\sqrt{1-4\tilde{K}\lambda}, \label{defg}
\end{eqnarray}
with $K(m)$ and $ E(m)$ as complete elliptic integrals of  the first
and second kind, respectively.
Consequently, the thermal boundary condition implies a switch in the usual features of the HH \cite{Hartle:1983ai} ($\epsilon=1$ choice in Eq.~(\ref{transition2})) and the tunnelling ($\epsilon=-1$ choice in Eq.~(\ref{transition2})) \cite{Vilenkin:1983xq,Linde:1983cm} 
wave functions: The  HH wave function,  once the  thermal boundary condition of \cite{Brustein:2005yn} is assumed, favours a non-vanishing cosmological constant; $\lambda\simeq 8.33\nu$, larger than  the one preferred by the tunnelling wave function; $\lambda\simeq 4\nu$, (see Fig.~1-b).

\vspace*{-0.3cm}
\begin{figure}[h]
\begin{center}
\includegraphics[width=4cm]{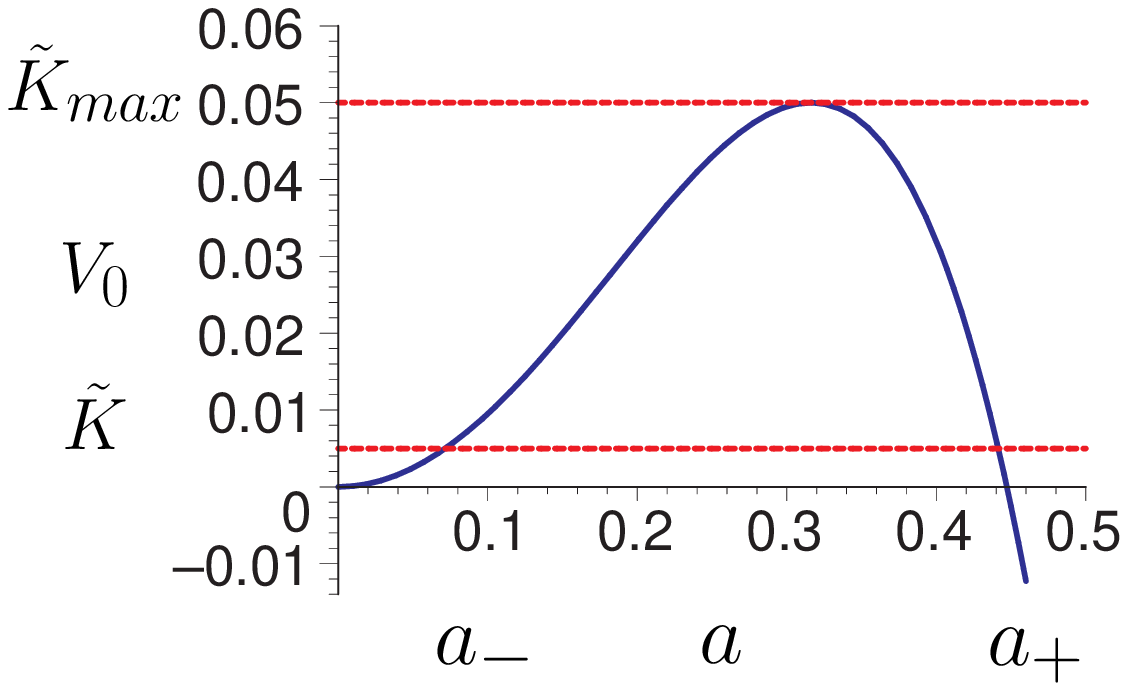}\hspace*{0.5cm}\includegraphics[width=4cm]{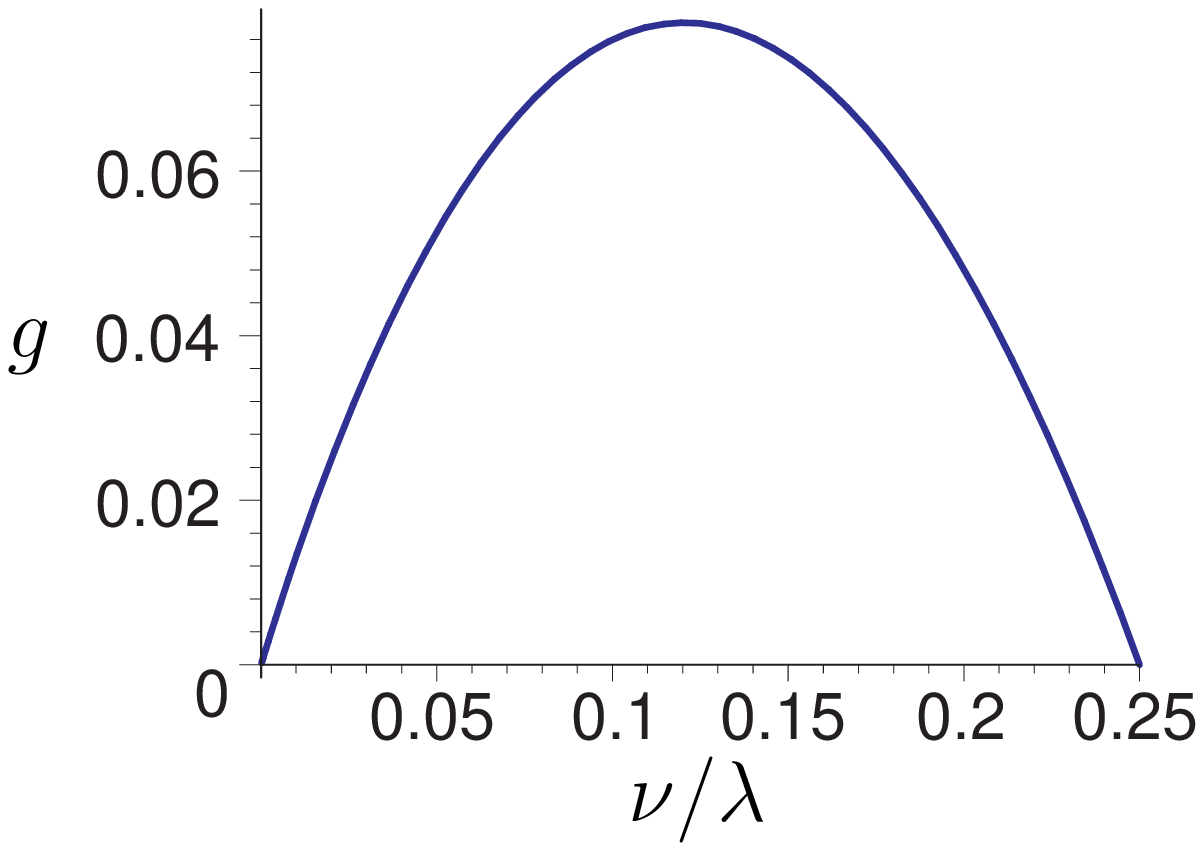}\hspace*{0.5cm}\includegraphics[width=4cm]{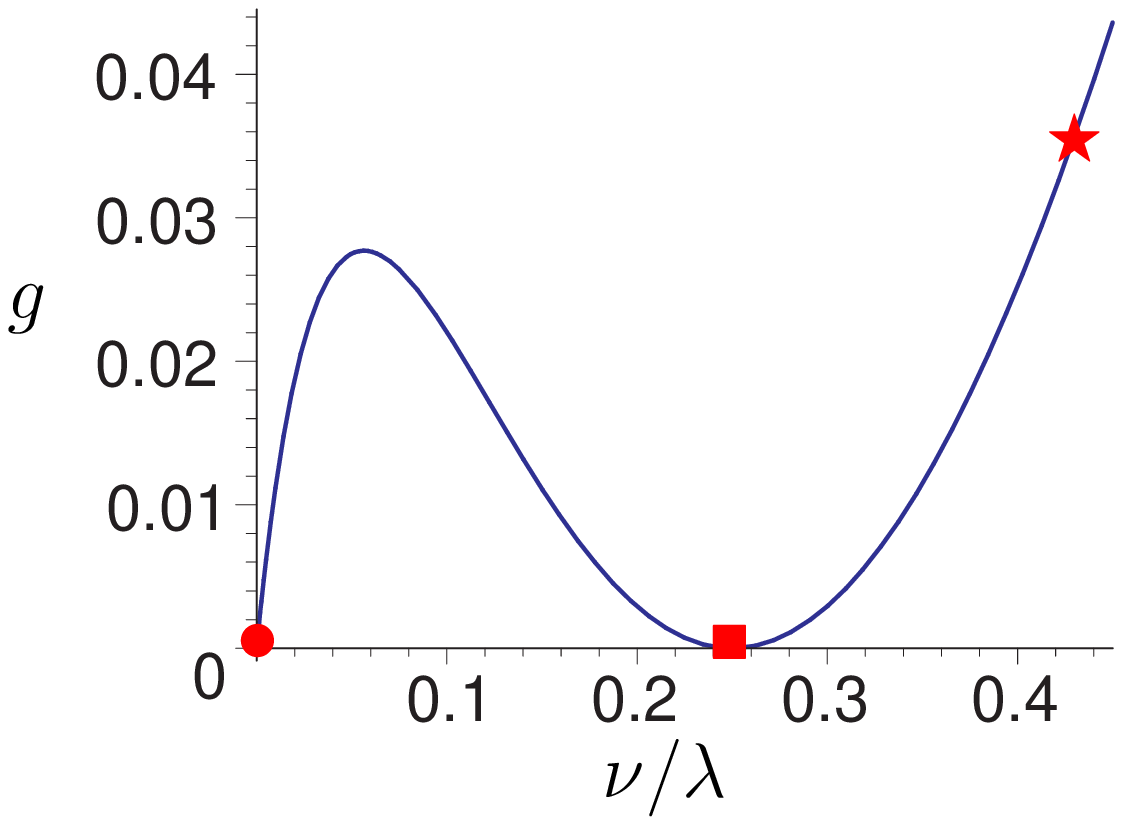}\\
\hspace*{0.7cm} 1-a \hspace*{3.8cm} 1-b \hspace*{4cm} 1-c
\end{center}
\vspace*{-0.3cm}
\caption{Figure 1-a corresponds to the potential barrier ($V_0-\tilde{K}$) separating the two Lorentzian regions of  a closed radiation-filled FRW Universe. $\tilde{K}_{\rm{max}}$ corresponds to the maximum amount of radiation consistent with the tunnelling of the Universe. We will refer to this situation as a large amount of radiation. Figures 1-b and 1-c corresponds to $g$ defined in Eq.~(\ref{defg}) for the thermal boundary condition and the {\em generalised thermal boundary condition}, respectively.} \label{amplitudeplot2}
\end{figure}

\vspace*{-0.5cm}

\enlargethispage*{8pt}

It turns out that the thermal effect considered in \cite{Brustein:2005yn} corresponds to a large amount of radiation where $\tilde{K}$ is close to $\tilde{K}_{\rm{max}}$ (see Fig.~1-a); i.e. the turning points $a_-$ and $a_+$ are very close or equivalently the height of the barrier separating both Lorentzian regions is very small.

Next we consider a {\em generalised
thermal boundary condition} for the wave function \cite{BLVM2006} where  we will assume instead an arbitrary amount of radiation, consistent with a  tunnelling
 of the Universe; i.e.  $\tilde{K}<\tilde{K}_{max}$ (see Fig.~1-a). Consequently, Eqs.~(\ref{K1})-(\ref{defg}) are replaced by consistent and more general relations where the amount of radiation as measured by $\tilde{K}$ depends also on $\lambda$ and reads
\begin{equation}
\tilde{K}= \frac{4\nu\lambda^{-2}}{(1+4\nu\lambda^{-1})^2}.
\label{eq20}\end{equation}
Within this broader range, the relevant features is that the transition amplitude as a function of $\nu/\lambda$ will be unlike the one deduced in \cite{Brustein:2005yn}. Indeed, this is the case as is shown in Figs.~1-b and 1-c. 

Regarding the HH wave function, it now favours a 
vanishing cosmological constant ($\nu /\lambda  
\rightarrow  \infty$) and $\tilde{K} \sim 1
/(4\nu)$. This physical case is represented 
schematically by a star in Fig.~1-c. In this manner, the role of the HH wave function 
and subsequent transition amplitude is returned to 
its ``original'' implication, with  the thermal  boundary 
condition being  implemented in a  fully consistent 
manner and not restricted to a narrow (perhaps not fully valid) 
limit. 

Concerning the tunnelling wave function, it  favours two possible physical situations depicted  by a circle and a square in Fig.~1-c. On the one hand, the ``circle'' option corresponds to a large cosmological constant ($\nu / \lambda \rightarrow 0$) and a small amount of radiation as measured by $\tilde{K}$ ($\tilde{K}
 \lambda \rightarrow 0$). On the other hand, the ``square'' option implies no tunnelling, that is,  $4 \nu / \lambda
\rightarrow 1$ or equivalently $4\tilde{K} \lambda \rightarrow 1$; i.e. both turning points coincide. In order to select 
one of these two possibilities for the tunnelling  wave function, we 
employed the DeWitt's argument \cite{DeWitt1,Bouhmadi-Lopez:2004mp}, since  
there is a curvature singularity at small scale factors. 
It turns out that the preferred value of the cosmological 
constant in this case is a large one. Moreover, this condition 
implies a small amount of radiation (as measured by the parameter 
$\tilde K$) allowing consequently the tunnelling of the Universe.  

\section{Conclusions}

We prove that the thermal boundary condition applied in  
\cite{Brustein:2005yn} corresponds to the particular physical situation 
where the amount of radiation is very large. We then provide a broader
and improved analysis of the {\em generalised thermal boundary condition} that is independent of such  restrictive limit \cite{BLVM2006}.

\enlargethispage*{8pt}

\vspace*{-0.3cm}

\section*{Acknowledgments}
MBL acknowledges the support of  a Marcel Grossmann fellowship to attend  the meeting.
MBL also acknowledges the support of CENTRA-IST BPD (Portugal) as well as the fellowship FCT/BPD/26542/2006.


\vspace*{-0.3cm}

\begin{thebibliography}{00}

\bibitem{Bousso:2000xa}
  R.~Bousso and J.~Polchinski,
  JHEP {\bf 0006}, 006 (2000). 
 

\bibitem{Douglas:2003um}
  M.~R.~Douglas,
  JHEP {\bf 0305}, 046 (2003).

\bibitem{Susskind:2003kw}
  L.~Susskind,
  arXiv:hep-th/0302219.

\bibitem{Hartle:1983ai}
  J.~B.~Hartle and S.~W.~Hawking,
  Phys.\ Rev.\ D {\bf 28}, 2960 (1983).

\bibitem{Vilenkin:1983xq}
  A.~Vilenkin,
  Phys.\ Rev.\ D {\bf 27}, 2848 (1983).

\bibitem{Linde:1983cm}
  A.~D.~Linde,
  Sov.\ Phys.\ JETP {\bf 60}, 211 (1984)
  [Zh.\ Eksp.\ Teor.\ Fiz.\  {\bf 87}, 369 (1984)].

\bibitem{Brustein:2005yn}
  R.~Brustein and S.~P.~de Alwis,
  Phys.\ Rev.\ D {\bf 73}, 046009 (2006).

\bibitem{Sarangi:2005cs}
  S.~Sarangi and S.~H.~Tye,
  arXiv:hep-th/0505104.

\bibitem{Holman:2005eu}
  R.~Holman and L.~Mersini-Houghton,
  Phys.\ Rev.\ D {\bf 74}, 043511 (2006).

\bibitem{BLVM2006}
 For more details see: M.~Bouhmadi-L\'{o}pez and P.~Vargas Moniz,
   arXiv:hep-th/0612149.

\bibitem{Vilenkin:1998rp}
  A.~Vilenkin,
  arXiv:gr-qc/9812027.

\bibitem{Bouhmadi-Lopez:2002qz}
  M.~Bouhmadi-L\'{o}pez, L.~J.~Garay and P.~F.~Gonz\'{a}lez-D\'{\i}az,
  Phys.\ Rev.\ D {\bf 66}, 083504 (2002).

\bibitem{DeWitt1}
  B.~DeWitt, Phys.\ Rev.\  {\bf 160}, 1113 (1967).

\bibitem{Bouhmadi-Lopez:2004mp}
  M.~Bouhmadi-L\'{o}pez and P.~Vargas Moniz,
  Phys.\ Rev.\ D {\bf 71}, 063521 (2005); 
  M.~Bouhmadi-Lopez and P.~Vargas Moniz,
  AIP Conf.\ Proc.\  {\bf 736}, 188 (2005).
\end{thebibliography}
\end{document}